# Optically Helicity-Dependent Orbital and Spin Dynamics in Two-Dimensional Ferromagnets

Shuo Li [1*], Ran Wang [1], Thomas Frauenheim [2], Junjie He [3*]

1 Institute for Advanced Study, Chengdu University, Chengdu 610106, China

2 Bremen Center for Computational Materials Science, University of Bremen, Bremen 28359, Germany

3 Department of Physical and Macromolecular Chemistry, Faculty of Science, Charles University in Prague, Prague 12843, Czech Republic

* Corresponding authors.

Email: shuoli.phd@gmail.com, junjie.he@natur.cuni.cz

## Abstract

Disentangling orbital (OAM) and spin (SAM) angular momenta in the ultrafast spin dynamics of two-dimensional (2D) ferromagnets on subfemtoseconds is a challenge in the field of ultrafast magnetism. Herein, we employed non-collinear spin version of real-time time-dependent density functional theory to investigate the orbital and spin dynamics of 2D ferromagnets $Fe_3GeTe_2$ (FGT) induced by circularly polarized light. Our results show the demagnetization of Fe sublattice in FGT is accompanied by helicity-dependent precession of OAM and SAM excited by circularly polarized lasers. We further identify that precession of OAM and SAM in FGT is faster than the demagnetization within a few femtoseconds. Remarkably, circularly polarized lasers can significantly induce a periodically transverse response of OAM and SAM on very ultrafast timescales of ~250 attoseconds. Our finding suggests a powerful new route for attosecond regimes of the angular momentum manipulation to coherently control helicity-dependent orbital and spin dynamics in 2D limits.

# Table of contents

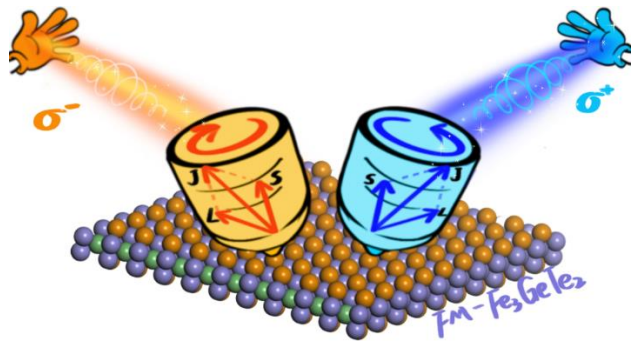

## Introduction

Ultrafast manipulation of the spin degree of freedom in magnetic materials serves as a fundamental principle for both storing and reading magnetic information by employing laser pulses[1-4]. These research not only enhances our knowledge of fundamental physics but also holds promise for developing faster, more efficient, and more reliable magnetic memory devices and exploring novel avenues for harnessing spin-related phenomena in future technological applications[5-8], with a primary focus on comprehending the underlying physical processes of ultrafast demagnetization and spin switching in magnets[9-11]. In two-dimensional (2D) magnetic materials[12, 13], a remarkably efficient exchange between the orbital and spin angular momentum can transpire due to the influence of spin-orbit coupling (SOC), originating from the relativistic motion of electrons. Distinguishing the individual contributions of the orbital and spin angular momentum in 2D limits presents a considerable challenge, primarily due to their rapid interplay occurring on a few femtoseconds[14-17].

The advent of femtoseconds and attosecond timescales enabled by ultrashort optical laser pulses offers new insights to ultrafast physics[14, 18]. By carefully studying the ultrafast dynamics induced by laser pulses, we can gain a deeper understanding of the transfer mechanisms and dynamics between orbital and spin degrees of freedom in 2D magnetic materials[13, 19]. All-optical helicity-dependent switching (AO-HDS)[20, 21] with the right circularly polarized and left circularly polarized pulses can switch magnetization without any external magnetic fields. The ultrafast circularly polarized lasers can switch the magnetization in 2D magnetic systems, such as $WSe_2$-$CrI_3$ van der Waals heterostructures[22]. However, the underlying mechanism of AO-HDS is still not conclusive. AO-HDS can be explained by the inverse Faraday effect[23-25] and the magnetic circular dichroism[26-28]. Moreover, the SOC and laser-induced superdiffusive spin currents in heterogeneous compounds may contribute for the AO-HDS process[29-32]. In fact, the microscopic perspective of AO-HDS is mainly derived from the optical spin transfer torque[33] and optical spin-orbit torque[34], which is determined by the interactions and orientations between laser pulses and spins. Although previous investigations exhibited some potential physical images for the dynamics of the angular momentum and demagnetization in AO-HDS process[31, 35, 36], the interactions between light, orbital and spin are much complicated and difficult to understand at a few femtoseconds. The microscopic mechanism of direct light-orbital-spin interactions so far is still not clear and needs to be explored further at a few femtoseconds.

The real-time time-dependent density functional theory (rt-TDDFT) can successfully describes the linearly polarized laser-induced demagnetization[37]. Dewhurst et al. [38] showed that the early spin dynamics based on the optical intersite spin transfer (OISTR) effect. This method supports us to investigate the spin dynamics and even the angular momentum in a few femtoseconds and even to distinguish the details on the attosecond timescales. Moreover, circularly polarized lasers have been triggered the generation of an orbital angular momentum in the nanoparticle[39] and explored the helicity-dependent spin dynamics of Fe, Co and Ni from the optical regime to the extreme ultraviolet regime.[40] Therefore, investigating orbital and spin dynamics and exploring the underlying mechanism of directly coherent light-orbital-spin interactions in 2D magnets will open new way the development of ultrafast spintronics driven by circularly polarized light.

2D ferromagnetic (FM) $Fe_3GeTe_2$ (FGT) [12, 41] enable us to explore light-matter interactions in the 2D limit, due to their interesting physical phenomena[42]. FGT could displays tunnelling magnetic Skyrmions[43], high spin-orbit torques[44], and strong electron correlation effects[45, 46]. Moreover, magnetic properties of FGT can be tuned by lase light. For example, Liu et al.[19] reported that the magnetic anisotropy of a few layered FGT can be manipulated by using a femtosecond laser pulse. The previous work showed that laser pulses induce significant large spin injection from FM FGT to nonmagnetic (NM) layers within a few femtoseconds, in which an interfacial atom-mediated spin transfer pathway in NM-FM heterostructures was identified[47]. Recently, Li et al.[48] reported that the intralayer spin transfer in graphene-FGT and silicence-FGT heterostructures could be controlled by Terahertz laser pulse. These novel physical features of FGT suggest that research on photoinduced the dynamic process in FGT could develop the new generation of ultrafast spintronics devices. However, the nature of the angular momentum flow and transfer between light, orbital and spin during ultrafast demagnetization phenomena remains elusive in the 2D regime. Therefore, the study of directly coherent interactions between light, orbit and spin on a few femtoseconds in FGT would be interesting and challenging.

In this work, we have investigated the orbital and spin dynamics in FGT excited by circularly polarized laser pulses using the non-collinear spin version of rt-TDDFT. The results show that the circularly polarized pump pulses can lead to different demagnetization processes due to the angular momentum flow and transfer between light, orbital and spin, which is deeper and more detailed than the commonly believed OISTR. Furthermore, we find that helicity-dependent demagnetization occurs on a few femtoseconds excited by circularly polarized laser pulses, and the angular momentum transfer between the angular momentum of polarized fields and the orbital angular momentum is faster than the demagnetization. This helicity-dependent orbital and spin dynamics would open a new way to manipulate the angular momentum for ultrafast spintronics.

## Demagnetization of FGT

The atomic structures of the optimized FGT monolayer are shown in Figure 1a. The lattice parameters of the unit cell of FGT are a = b = 4.051 Å. Three Fe atoms in the unit cell are located in two inequivalent sites, shown as $Fe_1/Fe_2$ and $Fe_3$. Each monolayer FGT consists of five sublayers. The FGT is metallic and the strong spin-orbital coupling (SOC) in Fe atoms lead to the out-plane magnetocrystalline anisotropy. The partial density of states (PDOS) of FGT is shown in Figure S1.

Itinerant magnets of Fe atoms can be demagnetized on femtoseconds excited by laser pulses[47]. We first simulated the spin dynamics of $M_z$ of each atom in FGT (Figure S2) excited by circularly left ($\sigma^+$) and right ($\sigma^-$) polarized lasers (Figure S3) under different frequencies (*f*). The simulations show the strong demagnetization processes of Fe atoms, which is mainly driven by photoinduced spin-selective charge transfer across both the majority and the minority states of Fe atoms in FGT[47, 48]. Moreover, the slight difference between $M_z^{\sigma^+}(t)$ and $M_z^{\sigma^-}(t)$ of the Fe atom (where $M_z^{\sigma^\pm}$ is the z component of the time-dependent magnetization for which dynamics is triggered by a $\sigma^+/\sigma^-$ polarized light) is derived from the helicity-dependent dynamics excited by circularly polarized lasers (Figure S4).

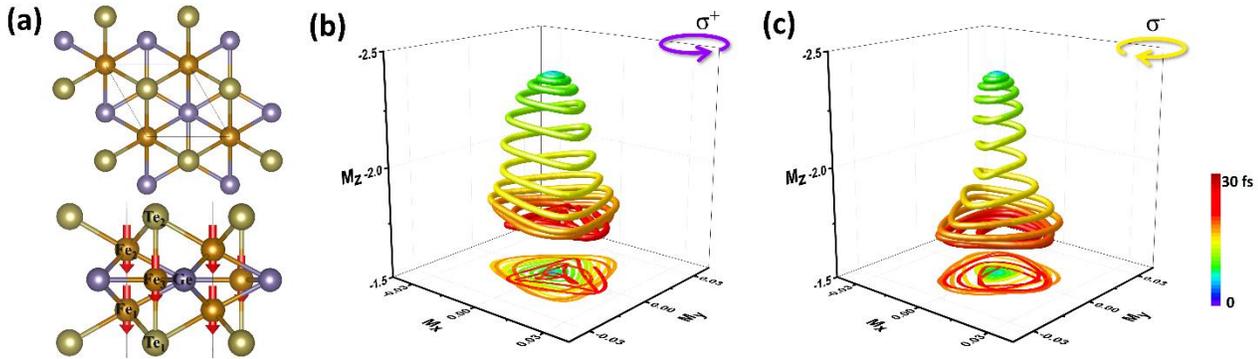

**Figure 1** (a) The atomic configuration of FGT. The red arrows located on Fe atoms represent the spin. The 3-dimensional demagnetization dynamics of the $Fe_1$ atom in FGT excited by circularly (b) left ($\sigma^+$) and (c) right ($\sigma^-$) polarized lasers. The color bar is the timescale from 0 to 30 fs.

## Transverse spin dynamics of FGT

The dynamics of the 3-dimensional magnetization are actually more complicated than its z-component, because the transverse components ($M_x$ and $M_y$) also change rapidly depending on the polarized lasers. Therefore, we show the 3-dimensional magnetization response of the $Fe_1$ atom to circularly $\sigma^+$ and $\sigma^-$ polarized lasers in 30 fs with F = 7.06 mJ/cm$^2$ and the full-width at half-maximum (FWHM) = 9.68 fs. The case of introducing the $\sigma^+$ polarized laser illustrates the $M_z$ quickly reducing and the anticlockwise rotation in the xy plane (Figure 1b). While the case of introducing the $\sigma^-$ polarized laser illustrates the $M_z$ quickly reducing and the clockwise rotation in the xy plane (Figure 1c). In addition, we have also calculated the 3-dimensional magnetization response of other atoms in FGT, in which the helicity-dependent magnetization processes of Te atoms are also shown (Figure S5), but not the middle layer of $Fe_3$ and Ge atoms. The results show that polarized lasers could induce the helicity-dependent demagnetization processes in FGT. Moreover, the dynamics of $M_x$ and $M_y$ exhibit the helicity-dependent of polarized lasers, indicating that the angular momentum could be transferred from the polarized fields of the laser pulses to the orbital and spin of atoms in FGT. Therefore, we will further distinguish the orbital and spin angular momentum in this system.

**Ultrafast response of the orbital angular momentum**

The light-matter interactions implies that the loss of the magnetic moments results in a transfer between the angular momentum of polarized fields of laser pulses, the orbital angular momentum (OAM, **L**) and the spin angular momentum (SAM, **S**). Taking the $Fe_1$ atom as an example, we can trigger different **L** and **S** in the $Fe_1$ atom of FGT excited by circularly polarized laser pulses. The angular momentum is only obtained when electrons are driven in two transverse axes at least. Our simulations show that the OAM in xy plane ($L_x$ and $L_y$) first increases and then decreases (Figure 2a), similar to circularly polarized laser pulses. Moreover, the anticlockwise and clockwise rotations of OAM in xy plane are observed by circularly $\sigma^+$ and $\sigma^-$ polarized lasers induced, respectively (Figure 2a). The SAM in xy plane ($S_x$ and $S_y$) also exhibits these anticlockwise and clockwise rotations (Figure 2b).

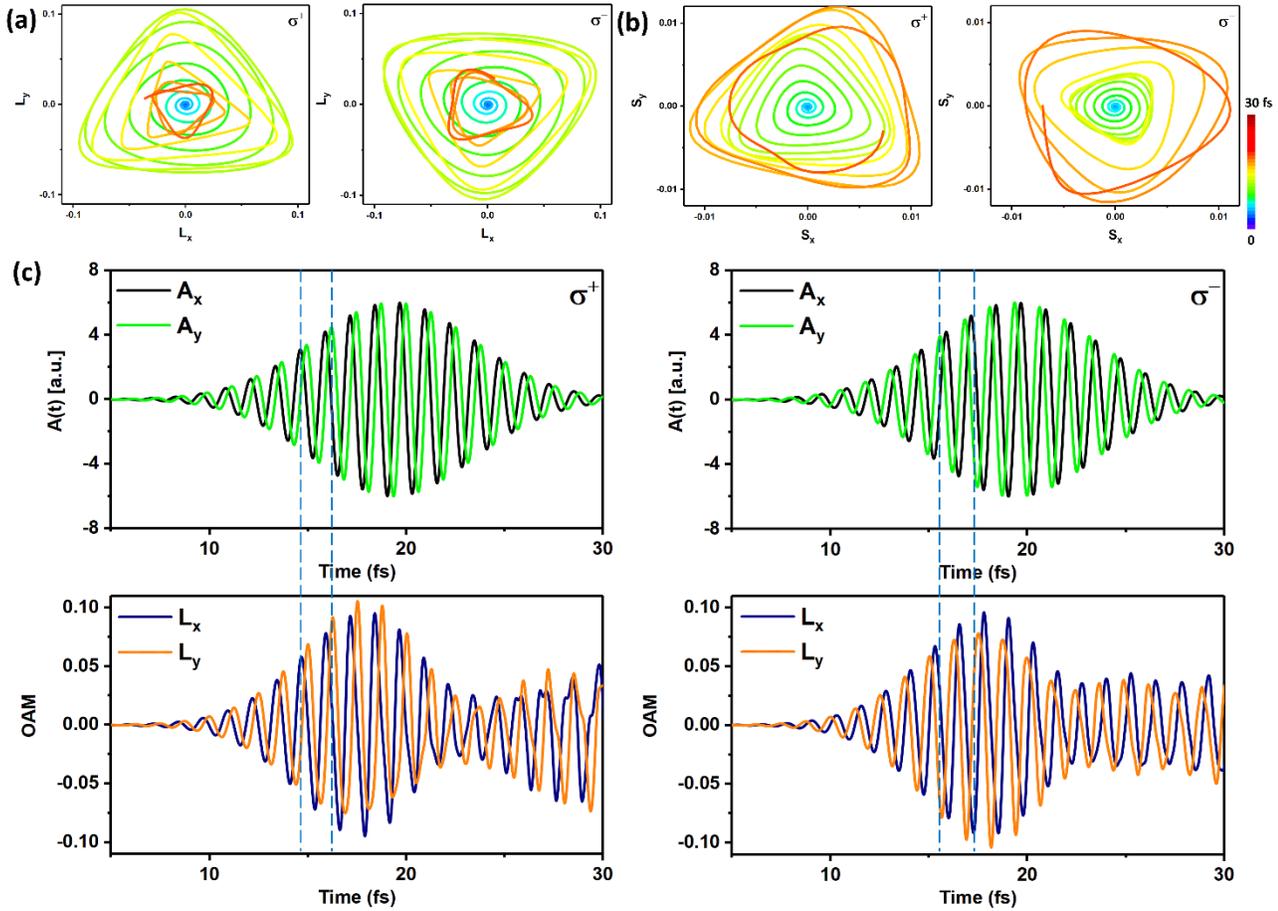

**Figure 2** The anticlockwise and clockwise rotations of (a) OAM in xy plane ($L_x$ and $L_y$) and (b) SAM in xy plane ($S_x$ and $S_y$) are shown excited by circularly $\sigma^+$ and $\sigma^-$ laser pulses, respectively. (c) Panels on the left and right columns represent the cases of circularly $\sigma^+$ and $\sigma^-$ polarized lasers induced, respectively. From top to bottom, the panels show the time dependence of the amplitude vector potential in xy plane ($A_x$ and $A_y$) of laser pulses and OAM.

To further understand the response between the angular momentum of polarized fields of laser pulses and **L**, we analyzed the angular momentum of polarized field and **L** along x and y orientation. Figure 3c shows that the on-resonance (left panels) and off-resonance (right panels) cases and exhibits several features of the polarized fields of laser pulses and the electronic response of **L** excited by circularly $\sigma^+$ and $\sigma^-$ laser pulses. The periodic response of on-resonance and off-resonance evolves intrinsically on very ultrafast timescales of ~250 attoseconds (from the extremum of $L_{x/y}$ to the next), based on the oscillation period of light, i.e. the frequency of the laser. Therefore, we can reasonably speculate that a higher frequency of circularly polarized lasers would lead to a faster response of the angular momentum. In addition, we tested different responses of **L** excited by circularly $\sigma^+$ and $\sigma^-$ laser pulses with different FWHM (Figure S6). The simulations show that the lifetime of the electronic response of **L** can keeps longer excited by circularly $\sigma^+$ and $\sigma^-$ laser pulses with a wide FWHM.

**Angular momentum transfer between orbital and spin**

An efficient transfer between orbital and spin magnetic moments in condensed matter can occur on femtoseconds due to the SOC effect[14]. Therefore, we now investigate the angular momentum transfer between orbital and spin in FGT after the ultrafast response of the orbital with circularly laser pulses. In the case of the $Fe_1$ atom excited by the circularly $\sigma^+$ laser pulse, the absolute value of $L_z$ first decreases and then increases (Figure 3a), indicating the direction of $L_z$ flips about 17.5 fs (Figure 3c) due to the opposite rotations between the polarized fields of laser pulses and the initial $L_z$. In the case of the $Fe_1$ atom excited by the circularly $\sigma^-$ laser pulse, the $L_z$ keeps in the -z direction (Figure 3a and 3c). However, the absolute value of $S_z$ decreases (Figure 3b) and their rotation angles ($\Delta\theta$) is less than 1 degree (Figure 4d). Interestingly, the $\Delta\theta$ of $L_z$ is much larger than that of $S_z$, indicating its rotational speed is slower than that of $S_z$, like a spinning top. The simulations show that the response between the polarized field of laser pulses and OAM (8.55 fs) is earlier than that of SAM (10.50 fs). Therefore, the response of **L** is faster than that of **S**, indicating that the angular momentum of polarized fields of laser pulses will first transfer to **L**, and thence to **S**. The hysteretic transfer between **L** and **S** could be from the SOC as a mediated step[39].

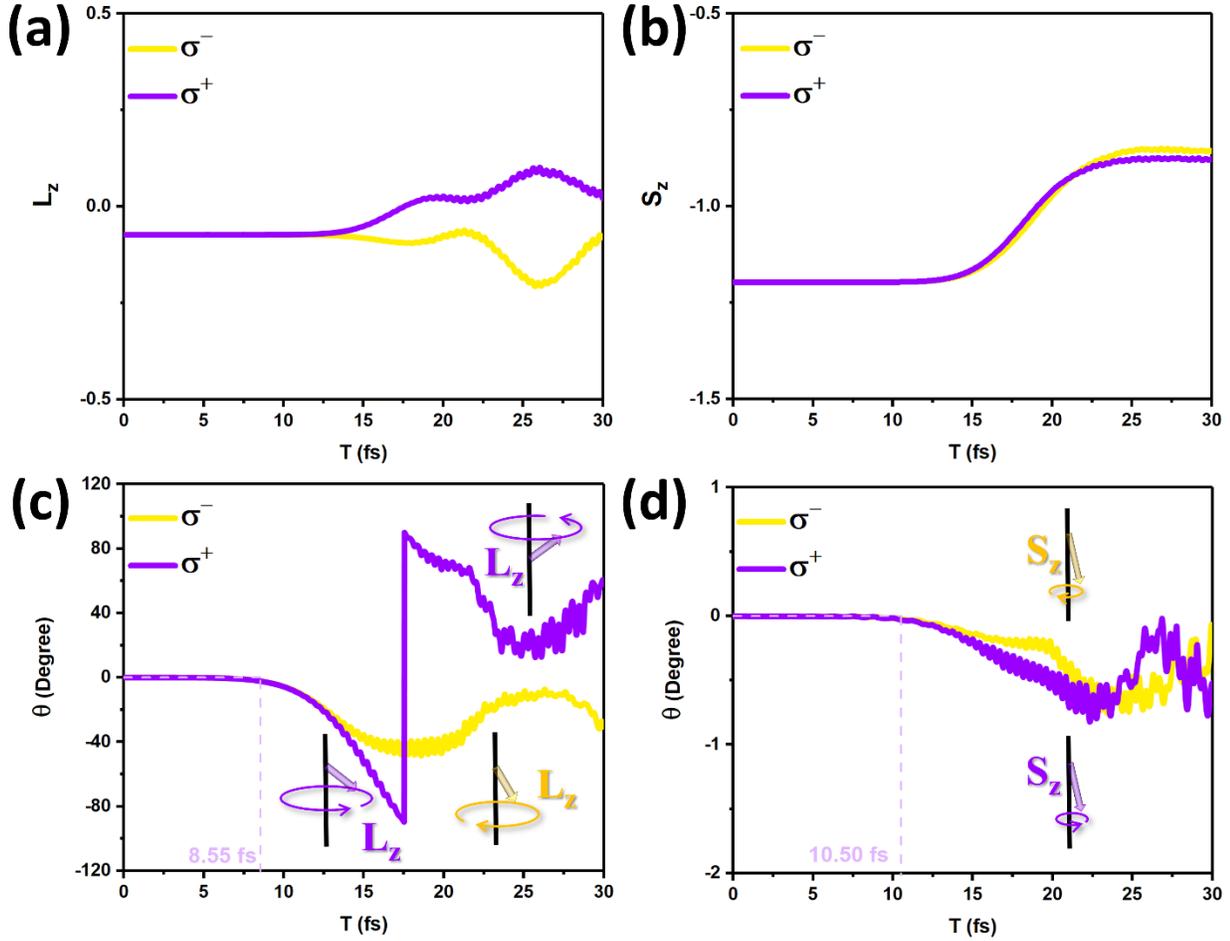

**Figure 3** (a) The dynamics of OAM along z direction ($L_z$) excited by circularly left ($\sigma^+$) and right ($\sigma^-$) polarized lasers. (b) The time evolution of $\Delta\theta$ for $L_z$. (c) The dynamics of SAM along z axis ($S_z$) excited by circularly left ($\sigma^+$) and right ($\sigma^-$) polarized laser pulses. (d) The time evolution of $\Delta\theta$ for $S_z$. The $\Delta\theta = \theta(t) - \theta(0)$ is the angles between the $L_z/S_z$ and z axis. The z direction is defined the positive direction. The initial time of the change of $\Delta\theta$ for $L_z$ and $S_z$ is marked with dotted lines and the precessional motion of $L_z$ and $S_z$ is shown in the illustration.

Based on the above analysis, we plotted the schematic diagram for the total angular momentum **J = L + S**, **L** and **S** excited by circularly polarized laser pulses (Figure 4a). Realization of these angular momentum transfer in a few femtoseconds would promise the new physical fundamental for ultrafast dynamics in 2D limits. To further distinguish the evolutions of J and demagnetization in FGT, we plotted the $\Delta\theta$ of J and the dynamics of $\Delta M$ excited by circularly $\sigma^+$ and $\sigma^-$ polarized lasers (Figure 5b), respectively. There is a delay of 4.11 fs between the $\theta$ of J and $\Delta M$, indicating that the angular momentum transfer between the polarized fields of laser pulses and J is faster than the demagnetization through the OISTR effect in FGT. Therefore, the manipulation of angular momentum using circularly polarized lasers could be the fastest methods currently available in 2D magnets for ultrafast spintronics.

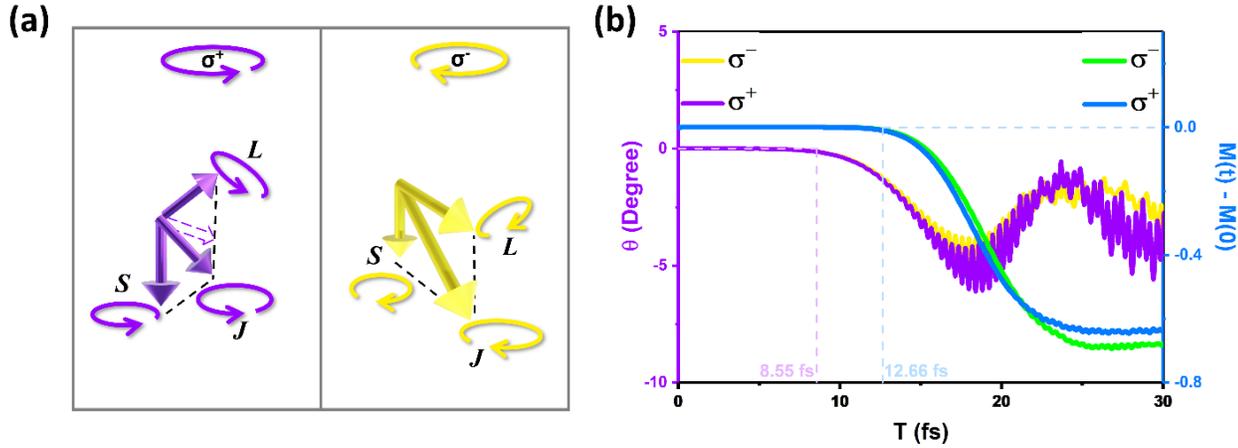

**Figure 4** (a) The schematic diagram for the evolution of **J**, **L** and **S** excited by circularly left ($\sigma^+$) and right ($\sigma^-$) polarized laser pulses. (b) The dynamics of $S_z$ excited by circularly polarized laser pulses. The magnetization $\Delta M = M(t) - M(0)$ as a function of time (in fs) for the $Fe_1$ atom of FGT. The initial time of the change of J and $\Delta M$ is marked with dotted lines and is shown in the illustration.

To further confirm the general regularity of the orbital and spin dynamics in 2D limits, we simulated the demagnetization of Cr atoms in two FM semiconductors of $Cr_2Ge_2Te_6$[49] (CGT) (Figure S7a) and $CrI_3$[50] (Figure S7b), which has experimentally prepared and exhibits the controllable magnetic properties including magnetic states and magnetic phase transitions by applying external fields[51-56]. The simulations show the 3-dimensional magnetization response of the Cr atom in CGT and $CrI_3$ by $\sigma^+$ and $\sigma^-$ polarized lasers in 30 fs. The anticlockwise and clockwise rotation demagnetization of Cr atoms in CGT (Figure S7c and S7d) and $CrI_3$ (Figure S7e and S7f) by $\sigma^+$ and $\sigma^-$ polarized lasers induced, respectively. Moreover, we also find that the periodic response between circularly polarized lasers and OAM of Cr sublayers in CGT (Figure S8) and $CrI_3$ (Figure S9). Therefore, the ultrafast orbital and spin dynamics in CGT and $CrI_3$ are similar to that of FGT. These results indicate that the manipulation of the orbital and spin angular momentum using circularly polarized lasers could be common and easy to achieve in 2D limits. This work will open up new research hotspots in the orbital and spin dynamics and will bring up new perspectives in ultrafast physics in the field of femtomagnetism and attomagnetism.

In conclusion, we demonstrate that for FGT, (i) the helicity-dependent orbital and spin dynamics not merely show at a few femtoseconds, not also they could be used to the ultrafast spintronics excited by circularly polarized laser pulses. (ii) The angular momentum transfer is faster than the demagnetization, and the periodic response of the angular momentum excited by circularly polarized laser pulses occurs in ~250 attoseconds. And (iii) the microscopic mechanism of the helicity-dependent orbital and spin dynamics is the ultrafast transfer between the angular momentum of polarized fields, OAM and SAM with the SOC effect, resulting in enhanced helicity-dependent demagnetization. These results pave the way for manipulating angular momentum by circularly polarized lasers in 2D magnets. The underlying mechanism of orbital and spin dynamics suggests that such 2D magnetic materials will provide a rich field for the manipulation of light-orbital-spin angular momentum transfer.

**Supporting Information**

The Supporting Information is available free of charge at DOI⋯

Methods and computational details, PDOS of FGT, demagnetization of FGT, circularly polarized fields with different frequencies, HI and HD spin dynamics, 3-dimensional demagnetization dynamics, OAM response.

**Acknowledgments**

We acknowledged the e-INFRA CZ (ID:90140) for providing computational resources were provided. We also acknowledged ChatGPT for improving readability and language. S. L. acknowledged the support from National Natural Science Foundation of China (Grant No. 12204069) and Natural Science Foundation of Sichuan Province of China (Grant No. 24NSFSC2537). J. H. acknowledged the support from MSCA Fellowships CZ – UK with CZ.02.01.01/00/22_010/0002902.

**Notes**

The authors declare that they have no competing interests.